\magnification=\magstep1
\font \authfont               = cmr10 scaled\magstep5
\font \fivesans               = cmss10 at 5pt
\font \headfont               = cmbx12 scaled\magstep4
\font \markfont               = cmr10 scaled\magstep1
\font \ninebf                 = cmbx9
\font \ninei                  = cmmi9
\font \nineit                 = cmti9
\font \ninerm                 = cmr9
\font \ninesans               = cmss10 at 9pt
\font \ninesl                 = cmsl9
\font \ninesy                 = cmsy9
\font \ninett                 = cmtt9
\font \sevensans              = cmss10 at 7pt
\font \sixbf                  = cmbx6
\font \sixi                   = cmmi6
\font \sixrm                  = cmr6
\font \sixsans                = cmss10 at 6pt
\font \sixsy                  = cmsy6
\font \smallescriptfont       = cmr5 at 7pt
\font \smallescriptscriptfont = cmr5
\font \smalletextfont         = cmr5 at 10pt
\font \subhfont               = cmr10 scaled\magstep4
\font \tafonts                = cmbx7  scaled\magstep2
\font \tafontss               = cmbx5  scaled\magstep2
\font \tafontt                = cmbx10 scaled\magstep2
\font \tams                   = cmmib10
\font \tamss                  = cmmib10 scaled 700
\font \tamt                   = cmmib10 scaled\magstep2
\font \tass                   = cmsy7  scaled\magstep2
\font \tasss                  = cmsy5  scaled\magstep2
\font \tast                   = cmsy10 scaled\magstep2
\font \tasys                  = cmex10 scaled\magstep1
\font \tasyt                  = cmex10 scaled\magstep2
\font \tbfonts                = cmbx7  scaled\magstep1
\font \tbfontss               = cmbx5  scaled\magstep1
\font \tbfontt                = cmbx10 scaled\magstep1
\font \tbms                   = cmmib10 scaled 833
\font \tbmss                  = cmmib10 scaled 600
\font \tbmt                   = cmmib10 scaled\magstep1
\font \tbss                   = cmsy7  scaled\magstep1
\font \tbsss                  = cmsy5  scaled\magstep1
\font \tbst                   = cmsy10 scaled\magstep1
\font \tenbfne                = cmb10
\font \tensans                = cmss10
\font \tpfonts                = cmbx7  scaled\magstep3
\font \tpfontss               = cmbx5  scaled\magstep3
\font \tpfontt                = cmbx10 scaled\magstep3
\font \tpmt                   = cmmib10 scaled\magstep3
\font \tpss                   = cmsy7  scaled\magstep3
\font \tpsss                  = cmsy5  scaled\magstep3
\font \tpst                   = cmsy10 scaled\magstep3
\font \tpsyt                  = cmex10 scaled\magstep3
\vsize=23true cm
\hsize=15true cm
\hfuzz=2pt
\tolerance=500
\abovedisplayskip=3 mm plus6pt minus 4pt
\belowdisplayskip=3 mm plus6pt minus 4pt
\abovedisplayshortskip=0mm plus6pt minus 2pt
\belowdisplayshortskip=2 mm plus4pt minus 4pt
\predisplaypenalty=0
\clubpenalty=10000
\widowpenalty=10000
\frenchspacing
\newdimen\oldparindent\oldparindent=1.5em
\parindent=1.5em
\skewchar\ninei='177 \skewchar\sixi='177
\skewchar\ninesy='60 \skewchar\sixsy='60
\hyphenchar\ninett=-1
\def\newline{\hfil\break}%
\catcode`@=11
\def\folio{\ifnum\pageno<\z@
\uppercase\expandafter{\romannumeral-\pageno}%
\else\number\pageno \fi}
\catcode`@=12 
  \mathchardef\Gamma="0100
  \mathchardef\Delta="0101
  \mathchardef\Theta="0102
  \mathchardef\Lambda="0103
  \mathchardef\Xi="0104
  \mathchardef\Pi="0105
  \mathchardef\Sigma="0106
  \mathchardef\Upsilon="0107
  \mathchardef\Phi="0108
  \mathchardef\Psi="0109
  \mathchardef\Omega="010A
  \mathchardef\bfGamma="0\the\bffam 00
  \mathchardef\bfDelta="0\the\bffam 01
  \mathchardef\bfTheta="0\the\bffam 02
  \mathchardef\bfLambda="0\the\bffam 03
  \mathchardef\bfXi="0\the\bffam 04
  \mathchardef\bfPi="0\the\bffam 05
  \mathchardef\bfSigma="0\the\bffam 06
  \mathchardef\bfUpsilon="0\the\bffam 07
  \mathchardef\bfPhi="0\the\bffam 08
  \mathchardef\bfPsi="0\the\bffam 09
  \mathchardef\bfOmega="0\the\bffam 0A

\def\sq{\hbox{\rlap{$\sqcap$}$\sqcup$}}

\def\utw{\smash{\rlap{\lower5pt\hbox{$\sim$}}}}
\def\udtw{\smash{\rlap{\lower6pt\hbox{$\approx$}}}}

\def\diameter{{\ifmmode\mathchoice
{\ooalign{\hfil\hbox{$\displaystyle/$}\hfil\crcr
{\hbox{$\displaystyle\mathchar"20D$}}}}
{\ooalign{\hfil\hbox{$\textstyle/$}\hfil\crcr
{\hbox{$\textstyle\mathchar"20D$}}}}
{\ooalign{\hfil\hbox{$\scriptstyle/$}\hfil\crcr
{\hbox{$\scriptstyle\mathchar"20D$}}}}
{\ooalign{\hfil\hbox{$\scriptscriptstyle/$}\hfil\crcr
{\hbox{$\scriptscriptstyle\mathchar"20D$}}}}
\else{\ooalign{\hfil/\hfil\crcr\mathhexbox20D}}%
\fi}}


\def\bbbc{{\mathchoice {\setbox0=\hbox{$\displaystyle\rm C$}\hbox{\hbox
to0pt{\kern0.4\wd0\vrule height0.9\ht0\hss}\box0}}
{\setbox0=\hbox{$\textstyle\rm C$}\hbox{\hbox
to0pt{\kern0.4\wd0\vrule height0.9\ht0\hss}\box0}}
{\setbox0=\hbox{$\scriptstyle\rm C$}\hbox{\hbox
to0pt{\kern0.4\wd0\vrule height0.9\ht0\hss}\box0}}
{\setbox0=\hbox{$\scriptscriptstyle\rm C$}\hbox{\hbox
to0pt{\kern0.4\wd0\vrule height0.9\ht0\hss}\box0}}}}
\def\bbbe{{\mathchoice {\setbox0=\hbox{\smalletextfont e}\hbox{\raise
0.1\ht0\hbox to0pt{\kern0.4\wd0\vrule width0.3pt height0.7\ht0\hss}\box0}}
{\setbox0=\hbox{\smalletextfont e}\hbox{\raise
0.1\ht0\hbox to0pt{\kern0.4\wd0\vrule width0.3pt height0.7\ht0\hss}\box0}}
{\setbox0=\hbox{\smallescriptfont e}\hbox{\raise
0.1\ht0\hbox to0pt{\kern0.5\wd0\vrule width0.2pt height0.7\ht0\hss}\box0}}
{\setbox0=\hbox{\smallescriptscriptfont e}\hbox{\raise
0.1\ht0\hbox to0pt{\kern0.4\wd0\vrule width0.2pt height0.7\ht0\hss}\box0}}}}
\def\bbbq{{\mathchoice {\setbox0=\hbox{$\displaystyle\rm Q$}\hbox{\raise
0.15\ht0\hbox to0pt{\kern0.4\wd0\vrule height0.8\ht0\hss}\box0}}
{\setbox0=\hbox{$\textstyle\rm Q$}\hbox{\raise
0.15\ht0\hbox to0pt{\kern0.4\wd0\vrule height0.8\ht0\hss}\box0}}
{\setbox0=\hbox{$\scriptstyle\rm Q$}\hbox{\raise
0.15\ht0\hbox to0pt{\kern0.4\wd0\vrule height0.7\ht0\hss}\box0}}
{\setbox0=\hbox{$\scriptscriptstyle\rm Q$}\hbox{\raise
0.15\ht0\hbox to0pt{\kern0.4\wd0\vrule height0.7\ht0\hss}\box0}}}}
\def\bbbt{{\mathchoice {\setbox0=\hbox{$\displaystyle\rm
T$}\hbox{\hbox to0pt{\kern0.3\wd0\vrule height0.9\ht0\hss}\box0}}
{\setbox0=\hbox{$\textstyle\rm T$}\hbox{\hbox
to0pt{\kern0.3\wd0\vrule height0.9\ht0\hss}\box0}}
{\setbox0=\hbox{$\scriptstyle\rm T$}\hbox{\hbox
to0pt{\kern0.3\wd0\vrule height0.9\ht0\hss}\box0}}
{\setbox0=\hbox{$\scriptscriptstyle\rm T$}\hbox{\hbox
to0pt{\kern0.3\wd0\vrule height0.9\ht0\hss}\box0}}}}
\def\bbbs{{\mathchoice
{\setbox0=\hbox{$\displaystyle     \rm S$}\hbox{\raise0.5\ht0\hbox
to0pt{\kern0.35\wd0\vrule height0.45\ht0\hss}\hbox
to0pt{\kern0.55\wd0\vrule height0.5\ht0\hss}\box0}}
{\setbox0=\hbox{$\textstyle        \rm S$}\hbox{\raise0.5\ht0\hbox
to0pt{\kern0.35\wd0\vrule height0.45\ht0\hss}\hbox
to0pt{\kern0.55\wd0\vrule height0.5\ht0\hss}\box0}}
{\setbox0=\hbox{$\scriptstyle      \rm S$}\hbox{\raise0.5\ht0\hbox
to0pt{\kern0.35\wd0\vrule height0.45\ht0\hss}\raise0.05\ht0\hbox
to0pt{\kern0.5\wd0\vrule height0.45\ht0\hss}\box0}}
{\setbox0=\hbox{$\scriptscriptstyle\rm S$}\hbox{\raise0.5\ht0\hbox
to0pt{\kern0.4\wd0\vrule height0.45\ht0\hss}\raise0.05\ht0\hbox
to0pt{\kern0.55\wd0\vrule height0.45\ht0\hss}\box0}}}}
\def\bbbz{{\mathchoice {\hbox{$\sans\textstyle Z\kern-0.4em Z$}}
{\hbox{$\sans\textstyle Z\kern-0.4em Z$}}
{\hbox{$\sans\scriptstyle Z\kern-0.3em Z$}}
{\hbox{$\sans\scriptscriptstyle Z\kern-0.2em Z$}}}}
\def\qed{\ifmmode\sq\else{\unskip\nobreak\hfil
\penalty50\hskip1em\null\nobreak\hfil\sq
\parfillskip=0pt\finalhyphendemerits=0\endgraf}\fi}
\newfam\sansfam
\textfont\sansfam=\tensans\scriptfont\sansfam=\sevensans
\scriptscriptfont\sansfam=\fivesans
\def\sans{\fam\sansfam\tensans}
\def\stackfigbox{\if
Y\FIG\global\setbox\figbox=\vbox{\unvbox\figbox\box1}%
\else\global\setbox\figbox=\vbox{\box1}\global\let\FIG=Y\fi}
\def\placefigure{\dimen0=\ht1\advance\dimen0by\dp1
\advance\dimen0by5\baselineskip
\advance\dimen0by0.4true cm
\ifdim\dimen0>\vsize\pageinsert\box1\vfill\endinsert
\else
\if Y\FIG\stackfigbox\else
\dimen0=\pagetotal\ifdim\dimen0<\pagegoal
\advance\dimen0by\ht1\advance\dimen0by\dp1\advance\dimen0by1.4true cm
\ifdim\dimen0>\pagegoal\stackfigbox
\else\box1\vskip4true mm\fi
\else\box1\vskip4true mm\fi\fi\fi}
%
\def\begfig#1cm#2\endfig{\par
\setbox1=\vbox{\dimen0=#1true cm\advance\dimen0
by1true cm\kern\dimen0#2}\placefigure}
\def\begdoublefig#1cm #2 #3 \enddoublefig{\begfig#1cm%
\vskip-.8333\baselineskip\line{\vtop{\hsize=0.46\hsize#2}\hfill
\vtop{\hsize=0.46\hsize#3}}\endfig}
\def\begfigsidebottom#1cm#2cm#3\endfigsidebottom{\dimen0=#2true cm
\ifdim\dimen0<0.4\hsize\message{Room for legend to narrow;
begfigsidebottom changed to begfig}\begfig#1cm#3\endfig\else
\par\def\figure##1##2{\vbox{\noindent\petit{\bf
Fig.\ts##1\unskip.\ }\ignorespaces ##2\par}}%
\dimen0=\hsize\advance\dimen0 by-.8true cm\advance\dimen0 by-#2true cm
\setbox1=\vbox{\hbox{\hbox to\dimen0{\vrule height#1true cm\hrulefill}%
\kern.8true cm\vbox{\hsize=#2true cm#3}}}\placefigure\fi}
\def\begfigsidetop#1cm#2cm#3\endfigsidetop{\dimen0=#2true cm
\ifdim\dimen0<0.4\hsize\message{Room for legend to narrow; begfigsidetop
changed to begfig}\begfig#1cm#3\endfig\else
\par\def\figure##1##2{\vbox{\noindent\petit{\bf
Fig.\ts##1\unskip.\ }\ignorespaces ##2\par}}%
\dimen0=\hsize\advance\dimen0 by-.8true cm\advance\dimen0 by-#2true cm
\setbox1=\vbox{\hbox{\hbox to\dimen0{\vrule height#1true cm\hrulefill}%
\kern.8true cm\vbox to#1true cm{\hsize=#2 true cm#3\vfill
}}}\placefigure\fi}
\def\figure#1#2{\vskip1true cm\setbox0=\vbox{\noindent\petit{\bf
Fig.\ts#1\unskip.\ }\ignorespaces #2\smallskip
\count255=0\global\advance\count255by\prevgraf}%
\ifnum\count255>1\box0\else
\centerline{\petit{\bf Fig.\ts#1\unskip.\
}\ignorespaces#2}\smallskip\fi}

\def\begtab#1cm#2\endtab{\par
   \ifvoid\topins\midinsert\medskip\vbox{#2\kern#1true cm}\endinsert
   \else\topinsert\vbox{#2\kern#1true cm}\endinsert\fi}
\def\begpet{\vskip6pt\bgroup\petit}
\def\endpet{\vskip6pt\egroup}
\newcount\frpages
\newcount\frpagegoal
\def\freepage#1{\global\frpagegoal=#1\relax\global\frpages=0\relax
\loop\global\advance\frpages by 1\relax
\message{Doing freepage \the\frpages\space of
\the\frpagegoal}\null\vfill\eject
\ifnum\frpagegoal>\frpages\repeat}
\newdimen\refindent
\def\begrefchapter#1{\titlea{}{\ignorespaces#1}%
\bgroup\petit
\setbox0=\hbox{1000.\enspace}\refindent=\wd0}
\def\ref{\goodbreak
\hangindent\oldparindent\hangafter=1
\noindent\ignorespaces}
\def\refno#1{\goodbreak
\hangindent\refindent\hangafter=1
\noindent\hbox to\refindent{#1\hss}\ignorespaces}
\def\endref{\goodbreak\endpet}
\def\vec#1{{\textfont1=\tams\scriptfont1=\tamss
\textfont0=\tenbf\scriptfont0=\sevenbf
\mathchoice{\hbox{$\displaystyle#1$}}{\hbox{$\textstyle#1$}}
{\hbox{$\scriptstyle#1$}}{\hbox{$\scriptscriptstyle#1$}}}}
\def\petit{\def\rm{\fam0\ninerm}%
\textfont0=\ninerm \scriptfont0=\sixrm \scriptscriptfont0=\fiverm
 \textfont1=\ninei \scriptfont1=\sixi \scriptscriptfont1=\fivei
 \textfont2=\ninesy \scriptfont2=\sixsy \scriptscriptfont2=\fivesy
 \def\it{\fam\itfam\nineit}%
 \textfont\itfam=\nineit
 \def\sl{\fam\slfam\ninesl}%
 \textfont\slfam=\ninesl
 \def\bf{\fam\bffam\ninebf}%
 \textfont\bffam=\ninebf \scriptfont\bffam=\sixbf
 \scriptscriptfont\bffam=\fivebf
 \def\sans{\fam\sansfam\ninesans}%
 \textfont\sansfam=\ninesans \scriptfont\sansfam=\sixsans
 \scriptscriptfont\sansfam=\fivesans
 \def\tt{\fam\ttfam\ninett}%
 \textfont\ttfam=\ninett
 \normalbaselineskip=11pt
 \setbox\strutbox=\hbox{\vrule height7pt depth2pt width0pt}%
 \normalbaselines\rm
\def\vec##1{{\textfont1=\tbms\scriptfont1=\tbmss
\textfont0=\ninebf\scriptfont0=\sixbf
\mathchoice{\hbox{$\displaystyle##1$}}{\hbox{$\textstyle##1$}}
{\hbox{$\scriptstyle##1$}}{\hbox{$\scriptscriptstyle##1$}}}}}
\nopagenumbers
%
\let\header=Y
\let\FIG=N
\newbox\figbox
\output={\if N\header\headline={\hfil}\fi\plainoutput\global\let\header=Y
\if Y\FIG\topinsert\unvbox\figbox\endinsert\global\let\FIG=N\fi}
\let\lasttitle=N
\def\bookauthor#1{\vfill\eject
     \bgroup
     \baselineskip=22pt
     \lineskip=0pt
     \pretolerance=10000
     \authfont
     \rightskip 0pt plus 6em
     \centerpar{#1}\vskip2true cm\egroup}
\def\bookhead#1#2{\bgroup
     \baselineskip=36pt
     \lineskip=0pt
     \pretolerance=10000
     \headfont
     \rightskip 0pt plus 6em
     \centerpar{#1}\vskip1true cm
     \baselineskip=22pt
     \subhfont\centerpar{#2}\vfill
     \parindent=0pt
     \baselineskip=16pt
     \leftskip=2.2true cm
     \markfont Springer-Verlag\newline
     Berlin Heidelberg New York\newline
     London Paris Tokyo Singapore\bigskip\bigskip
     [{\it This is page III of your manuscript and will be reset by
     Springer.}]
     \egroup\let\header=N\eject}
\def\centerpar#1{{\parfillskip=0pt
\rightskip=0pt plus 1fil
\leftskip=0pt plus 1fil
\advance\leftskip by\oldparindent
\advance\rightskip by\oldparindent
\def\newline{\break}%
\noindent\ignorespaces#1\par}}
\def\part#1#2{\vfill\supereject\let\header=N
\centerline{\subhfont#1}%
\vskip75pt
     \bgroup
\textfont0=\tpfontt \scriptfont0=\tpfonts \scriptscriptfont0=\tpfontss
\textfont1=\tpmt \scriptfont1=\tbmt \scriptscriptfont1=\tams
\textfont2=\tpst \scriptfont2=\tpss \scriptscriptfont2=\tpsss
\textfont3=\tpsyt \scriptfont3=\tasys \scriptscriptfont3=\tenex
     \baselineskip=20pt
     \lineskip=0pt
     \pretolerance=10000
     \tpfontt
     \centerpar{#2}
     \vfill\eject\egroup\ignorespaces}
\newtoks\AUTHOR
\newtoks\HEAD
\catcode`\@=\active
\def\author#1{\bgroup
\baselineskip=22pt
\lineskip=0pt
\pretolerance=10000
\markfont
\centerpar{#1}\bigskip\egroup
{\def@##1{}%
\setbox0=\hbox{\petit\kern2.5true cc\ignorespaces#1\unskip}%
\ifdim\wd0>\hsize
\message{The names of the authors exceed the headline, please use a }%
\message{short form with AUTHORRUNNING}\gdef\leftheadline{%
\hbox to2.5true cc{\folio\hfil}AUTHORS suppressed due to excessive
length\hfil}%
\global\AUTHOR={AUTHORS were to long}\else
\xdef\leftheadline{\hbox to2.5true
cc{\noexpand\folio\hfil}\ignorespaces#1\hfill}%
\global\AUTHOR={\def@##1{}\ignorespaces#1\unskip}\fi
}\let\INS=E}
\def\address#1{\bgroup
\centerpar{#1}\bigskip\egroup
\catcode`\@=12
\vskip2cm\noindent\ignorespaces}
\let\INS=N%
\def@#1{\if N\INS\unskip\ $^{#1}$\else\if
E\INS\noindent$^{#1}$\let\INS=Y\ignorespaces
\else\par
\noindent$^{#1}$\ignorespaces\fi\fi}%
\catcode`\@=12
\headline={\petit\def\newline{ }\def\fonote#1{}\ifodd\pageno
\rightheadline\else\leftheadline\fi}
\def\rightheadline{\hfil Missing CONTRIBUTION
title\hbox to2.5true cc{\hfil\folio}}
\def\leftheadline{\hbox to2.5true cc{\folio\hfil}Missing name(s) of the
author(s)\hfil}
\nopagenumbers
\let\header=Y

\let\lasttitle=N
 \def\contribution#1{\vfill\supereject
 \ifodd\pageno\else\null\vfill\supereject\fi
 \let\header=N\bgroup
 \textfont0=\tafontt \scriptfont0=\tafonts \scriptscriptfont0=\tafontss
 \textfont1=\tamt \scriptfont1=\tams \scriptscriptfont1=\tams
 \textfont2=\tast \scriptfont2=\tass \scriptscriptfont2=\tasss
 \par\baselineskip=16pt
     \lineskip=16pt
     \tafontt
     \raggedright
     \pretolerance=10000
     \noindent
     \centerpar{\ignorespaces#1}%
     \vskip12pt\egroup
     \nobreak
     \parindent=0pt
     \everypar={\global\parindent=1.5em
     \global\let\lasttitle=N\global\everypar={}}%
     \global\let\lasttitle=A%
     \setbox0=\hbox{\petit\def\newline{ }\def\fonote##1{}\kern2.5true
     cc\ignorespaces#1}\ifdim\wd0>\hsize
     \message{Your CONTRIBUTIONtitle exceeds the headline,
please use a short form
with CONTRIBUTIONRUNNING}\gdef\rightheadline{\hfil CONTRIBUTION title
suppressed due to excessive length\hbox to2.5true cc{\hfil\folio}}%
\global\HEAD={HEAD was to long}\else
\gdef\rightheadline{\hfill\ignorespaces#1\unskip\hbox to2.5true
cc{\hfil\folio}}\global\HEAD={\ignorespaces#1\unskip}\fi
\catcode`\@=\active
     \ignorespaces}
 \def\contributionnext#1{\vfill\supereject
 \let\header=N\bgroup
 \textfont0=\tafontt \scriptfont0=\tafonts \scriptscriptfont0=\tafontss
 \textfont1=\tamt \scriptfont1=\tams \scriptscriptfont1=\tams
 \textfont2=\tast \scriptfont2=\tass \scriptscriptfont2=\tasss
 \par\baselineskip=16pt
     \lineskip=16pt
     \tafontt
     \raggedright
     \pretolerance=10000
     \noindent
     \centerpar{\ignorespaces#1}%
     \vskip12pt\egroup
     \nobreak
     \parindent=0pt
     \everypar={\global\parindent=1.5em
     \global\let\lasttitle=N\global\everypar={}}%
     \global\let\lasttitle=A%
     \setbox0=\hbox{\petit\def\newline{ }\def\fonote##1{}\kern2.5true
     cc\ignorespaces#1}\ifdim\wd0>\hsize
     \message{Your CONTRIBUTIONtitle exceeds the headline,
please use a short form
with CONTRIBUTIONRUNNING}\gdef\rightheadline{\hfil CONTRIBUTION title
suppressed due to excessive length\hbox to2.5true cc{\hfil\folio}}%
\global\HEAD={HEAD was to long}\else
\gdef\rightheadline{\hfill\ignorespaces#1\unskip\hbox to2.5true
cc{\hfil\folio}}\global\HEAD={\ignorespaces#1\unskip}\fi
\catcode`\@=\active
     \ignorespaces}
\def\motto#1#2{\bgroup\petit\leftskip=6.5true cm\noindent\ignorespaces#1
\if!#2!\else\medskip\noindent\it\ignorespaces#2\fi\bigskip\egroup
\let\lasttitle=M
\parindent=0pt
\everypar={\global\parindent=\oldparindent
\global\let\lasttitle=N\global\everypar={}}%
\global\let\lasttitle=M%
\ignorespaces}
\def\abstract#1{\bgroup\petit\noindent
{\bf Abstract: }\ignorespaces#1\vskip28pt\egroup
\let\lasttitle=N
\parindent=0pt
\everypar={\global\parindent=\oldparindent
\global\let\lasttitle=N\global\everypar={}}%
\ignorespaces}
\def\titlea#1#2{\if N\lasttitle\else\vskip-28pt
     \fi
     \vskip18pt plus 4pt minus4pt
     \bgroup
\textfont0=\tbfontt \scriptfont0=\tbfonts \scriptscriptfont0=\tbfontss
\textfont1=\tbmt \scriptfont1=\tbms \scriptscriptfont1=\tbmss
\textfont2=\tbst \scriptfont2=\tbss \scriptscriptfont2=\tbsss
\textfont3=\tasys \scriptfont3=\tenex \scriptscriptfont3=\tenex
     \baselineskip=16pt
     \lineskip=0pt
     \pretolerance=10000
     \noindent
     \tbfontt
     \rightskip 0pt plus 6em
     \setbox0=\vbox{\vskip23pt\def\fonote##1{}%
     \noindent
     \if!#1!\ignorespaces#2
     \else\setbox0=\hbox{\ignorespaces#1\unskip\ }\hangindent=\wd0
     \hangafter=1\box0\ignorespaces#2\fi
     \vskip18pt}%
     \dimen0=\pagetotal\advance\dimen0 by-\pageshrink
     \ifdim\dimen0<\pagegoal
     \dimen0=\ht0\advance\dimen0 by\dp0\advance\dimen0 by
     3\normalbaselineskip
     \advance\dimen0 by\pagetotal
     \ifdim\dimen0>\pagegoal\eject\fi\fi
     \noindent
     \if!#1!\ignorespaces#2
     \else\setbox0=\hbox{\ignorespaces#1\unskip\ }\hangindent=\wd0
     \hangafter=1\box0\ignorespaces#2\fi
     \vskip18pt plus4pt minus4pt\egroup
     \nobreak
     \parindent=0pt
     \everypar={\global\parindent=\oldparindent
     \global\let\lasttitle=N\global\everypar={}}%
     \global\let\lasttitle=A%
     \ignorespaces}
 \def\titleb#1#2{\if N\lasttitle\else\vskip-28pt
     \fi
     \vskip18pt plus 4pt minus4pt
     \bgroup
\textfont0=\tenbf \scriptfont0=\sevenbf \scriptscriptfont0=\fivebf
\textfont1=\tams \scriptfont1=\tamss \scriptscriptfont1=\tbmss
     \lineskip=0pt
     \pretolerance=10000
     \noindent
     \bf
     \rightskip 0pt plus 6em
     \setbox0=\vbox{\vskip23pt\def\fonote##1{}%
     \noindent
     \if!#1!\ignorespaces#2
     \else\setbox0=\hbox{\ignorespaces#1\unskip\enspace}\hangindent=\wd0
     \hangafter=1\box0\ignorespaces#2\fi
     \vskip10pt}%
     \dimen0=\pagetotal\advance\dimen0 by-\pageshrink
     \ifdim\dimen0<\pagegoal
     \dimen0=\ht0\advance\dimen0 by\dp0\advance\dimen0 by
     3\normalbaselineskip
     \advance\dimen0 by\pagetotal
     \ifdim\dimen0>\pagegoal\eject\fi\fi
     \noindent
     \if!#1!\ignorespaces#2
     \else\setbox0=\hbox{\ignorespaces#1\unskip\enspace}\hangindent=\wd0
     \hangafter=1\box0\ignorespaces#2\fi
     \vskip8pt plus4pt minus4pt\egroup
     \nobreak
     \parindent=0pt
     \everypar={\global\parindent=\oldparindent
     \global\let\lasttitle=N\global\everypar={}}%
     \global\let\lasttitle=B%
     \ignorespaces}
 \def\titlec#1#2{\if N\lasttitle\else\vskip-23pt
     \fi
     \vskip18pt plus 4pt minus4pt
     \bgroup
\textfont0=\tenbfne \scriptfont0=\sevenbf \scriptscriptfont0=\fivebf
\textfont1=\tams \scriptfont1=\tamss \scriptscriptfont1=\tbmss
     \tenbfne
     \lineskip=0pt
     \pretolerance=10000
     \noindent
     \rightskip 0pt plus 6em
     \setbox0=\vbox{\vskip23pt\def\fonote##1{}%
     \noindent
     \if!#1!\ignorespaces#2
     \else\setbox0=\hbox{\ignorespaces#1\unskip\enspace}\hangindent=\wd0
     \hangafter=1\box0\ignorespaces#2\fi
     \vskip6pt}%
     \dimen0=\pagetotal\advance\dimen0 by-\pageshrink
     \ifdim\dimen0<\pagegoal
     \dimen0=\ht0\advance\dimen0 by\dp0\advance\dimen0 by
     2\normalbaselineskip
     \advance\dimen0 by\pagetotal
     \ifdim\dimen0>\pagegoal\eject\fi\fi
     \noindent
     \if!#1!\ignorespaces#2
     \else\setbox0=\hbox{\ignorespaces#1\unskip\enspace}\hangindent=\wd0
     \hangafter=1\box0\ignorespaces#2\fi
     \vskip6pt plus4pt minus4pt\egroup
     \nobreak
     \parindent=0pt
     \everypar={\global\parindent=\oldparindent
     \global\let\lasttitle=N\global\everypar={}}%
     \global\let\lasttitle=C%
     \ignorespaces}
 \def\titled#1{\if N\lasttitle\else\vskip-\baselineskip
     \fi
     \vskip12pt plus 4pt minus 4pt
     \bgroup
\textfont1=\tams \scriptfont1=\tamss \scriptscriptfont1=\tbmss
     \bf
     \noindent
     \ignorespaces#1\ \ignorespaces\egroup
     \ignorespaces}
\let\ts=\thinspace
\def\footnoterule{\kern-3pt\hrule width 2true cm\kern2.6pt}
\newcount\footcount \footcount=0
\def\advftncnt{\advance\footcount by1\global\footcount=\footcount}
\def\fonote#1{\advftncnt$^{\the\footcount}$\begingroup\petit
\parfillskip=0pt plus 1fil
\def\textindent##1{\hangindent0.5\oldparindent\noindent\hbox
to0.5\oldparindent{##1\hss}\ignorespaces}%
\vfootnote{$^{\the\footcount}$}{#1\vskip-9.69pt}\endgroup}
\def\item#1{\par\noindent
\hangindent6.5 mm\hangafter=0
\llap{#1\enspace}\ignorespaces}

\def\titleao#1{\vfill\supereject
\ifodd\pageno\else\null\vfill\supereject\fi
\let\header=N
     \bgroup
\textfont0=\tafontt \scriptfont0=\tafonts \scriptscriptfont0=\tafontss
\textfont1=\tamt \scriptfont1=\tams \scriptscriptfont1=\tamss
\textfont2=\tast \scriptfont2=\tass \scriptscriptfont2=\tasss
\textfont3=\tasyt \scriptfont3=\tasys \scriptscriptfont3=\tenex
     \baselineskip=18pt
     \lineskip=0pt
     \pretolerance=10000
     \tafontt
     \centerpar{#1}%
     \vskip75pt\egroup
     \nobreak
     \parindent=0pt
     \everypar={\global\parindent=\oldparindent
     \global\let\lasttitle=N\global\everypar={}}%
     \global\let\lasttitle=A%
     \ignorespaces}






\def\leaderfill{\kern0.5em\leaders\hbox to 0.5em{\hss.\hss}\hfill\kern
0.5em}
\newdimen\chapindent
\newdimen\sectindent
\newdimen\subsecindent
\newdimen\thousand
\setbox0=\hbox{\bf 10. }\chapindent=\wd0
\setbox0=\hbox{10.10 }\sectindent=\wd0
\setbox0=\hbox{10.10.1 }\subsecindent=\wd0
\setbox0=\hbox{\enspace 100}\thousand=\wd0
\def\contpart#1#2{\medskip\noindent
\vbox{\kern10pt\leftline{\textfont1=\tams
\scriptfont1=\tamss\scriptscriptfont1=\tbmss\bf
\advance\chapindent by\sectindent
\hbox to\chapindent{\ignorespaces#1\hss}\ignorespaces#2}\kern8pt}%
\let\lasttitle=Y\par}
\def\contcontribution#1#2{\if N\lasttitle\bigskip\fi
\let\lasttitle=N\line{{\textfont1=\tams
\scriptfont1=\tamss\scriptscriptfont1=\tbmss\bf#1}%
\if!#2!\hfill\else\leaderfill\hbox to\thousand{\hss#2}\fi}\par}
\def\conttitlea#1#2#3{\line{\hbox to
\chapindent{\strut\bf#1\hss}{\textfont1=\tams
\scriptfont1=\tamss\scriptscriptfont1=\tbmss\bf#2}%
\if!#3!\hfill\else\leaderfill\hbox to\thousand{\hss#3}\fi}\par}
\def\conttitleb#1#2#3{\line{\kern\chapindent\hbox
to\sectindent{\strut#1\hss}{#2}%
\if!#3!\hfill\else\leaderfill\hbox to\thousand{\hss#3}\fi}\par}
\def\conttitlec#1#2#3{\line{\kern\chapindent\kern\sectindent
\hbox to\subsecindent{\strut#1\hss}{#2}%
\if!#3!\hfill\else\leaderfill\hbox to\thousand{\hss#3}\fi}\par}
\long\def\lemma#1#2{\removelastskip\vskip\baselineskip\noindent{\tenbfne
Lemma\if!#1!\else\ #1\fi\ \ }{\it\ignorespaces#2}\vskip\baselineskip}
\long\def\proposition#1#2{\removelastskip\vskip\baselineskip\noindent{\tenbfne
Proposition\if!#1!\else\ #1\fi\ \ }{\it\ignorespaces#2}\vskip\baselineskip}
\long\def\theorem#1#2{\removelastskip\vskip\baselineskip\noindent{\tenbfne
Theorem\if!#1!\else\ #1\fi\ \ }{\it\ignorespaces#2}\vskip\baselineskip}
\long\def\corollary#1#2{\removelastskip\vskip\baselineskip\noindent{\tenbfne
Corollary\if!#1!\else\ #1\fi\ \ }{\it\ignorespaces#2}\vskip\baselineskip}
\long\def\example#1#2{\removelastskip\vskip\baselineskip\noindent{\tenbfne
Example\if!#1!\else\ #1\fi\ \ }\ignorespaces#2\vskip\baselineskip}
\long\def\exercise#1#2{\removelastskip\vskip\baselineskip\noindent{\tenbfne
Exercise\if!#1!\else\ #1\fi\ \ }\ignorespaces#2\vskip\baselineskip}
\long\def\problem#1#2{\removelastskip\vskip\baselineskip\noindent{\tenbfne
Problem\if!#1!\else\ #1\fi\ \ }\ignorespaces#2\vskip\baselineskip}
\long\def\solution#1#2{\removelastskip\vskip\baselineskip\noindent{\tenbfne
Solution\if!#1!\else\ #1\fi\ \ }\ignorespaces#2\vskip\baselineskip}


\long\def\definition#1#2{\removelastskip\vskip\baselineskip\noindent{\tenbfne
Definition\if!#1!\else\
#1\fi\ \ }\ignorespaces#2\vskip\baselineskip}
\def\frame#1{\bigskip\vbox{\hrule\hbox{\vrule\kern5pt
\vbox{\kern5pt\advance\hsize by-10.8pt
\centerline{\vbox{#1}}\kern5pt}\kern5pt\vrule}\hrule}\bigskip}
\def\frameddisplay#1#2{$$\vcenter{\hrule\hbox{\vrule\kern5pt
\vbox{\kern5pt\hbox{$\displaystyle#1$}%
\kern5pt}\kern5pt\vrule}\hrule}\eqno#2$$}
\def\typeset{\petit\noindent This book was processed by the author using
the \TeX\ macro package from Springer-Verlag.\par}
\outer\def\byebye{\bigskip\bigskip\typeset
\footcount=1\ifx\speciali\undefined\else
\loop\smallskip\noindent special character No\number\footcount:
\csname special\romannumeral\footcount\endcsname
\advance\footcount by 1\global\footcount=\footcount
\ifnum\footcount<11\repeat\fi
\gdef\leftheadline{\hbox to2.5true cc{\folio\hfil}\ignorespaces
\the\AUTHOR\unskip: \the\HEAD\hfill}\vfill\supereject\end}

\topinsert
\vbox{\hrule\smallskip\centerline{\hbox{Talk presented at the conference ``Relativistic Jets
in AGN'',}} \centerline{\hbox{Cracow, May 1997, M. Ostrowski, M. Sikora, G. Madejski, M. Begelman (eds.)}}\smallskip\hrule}\endinsert
\def\atsign{\hbox{@}}
\contribution{Relativistic Jets in Radio-Weak Quasars and LINER Galaxies}
\author{@{1,2}Heino Falcke, @1Andrew S. Wilson, @3Luis C. Ho}
\address{@1Department of Astronomy, University of Maryland
       College Park, MD 20742-2421, USA
@2Max-Planck-institut f\"ur Radioastronomie, Auf dem H\"ugel 69, D-53121 Bonn, Germany
(hfalcke\atsign{}mpifr-bonn.mpg.de)
@3Harvard-Smithsonian Center for Astrophysics, Cambridge, MA 02138, USA
}
\abstract{
We discuss radio observations of a sample of radio-weak and
radio-inter\-mediate quasars which demonstrate that -- just like their
radio-loud counterparts -- radio-weak quasars too have relativistic
jets in their nuclei. Moreover, a VLA survey of nearby LINER galaxies
reveals a relatively large number of flat-spectrum radio cores. It is
suggested that those cores are also best explained by moderately
relativistic jets ($\gamma\sim2$) produced by a central
engine.}

\titlea{1}{Introduction}
When we discuss the properties of relativistic jets in AGN, we usually
tend to think about radio galaxies and radio-loud, core-dominated
quasars. The observation of superluminal motion and of many other
indicators of high bulk Lorentz factors in these sources have
established the existence of relativistic jets there beyond any doubt.
But is this the whole universe, or just the tip of the iceberg?

In comparison to stellar winds it is often argued that the escape
speed from the central object is an important factor that determines
the terminal jet speed. If that is true and since we believe that most
of the AGN are powered by a black hole --- which in fact may
eventually become a definition rather than a conclusion --- one should
assume that if an AGN produces a jet it should {\it always} be
relativistic. Consequently the crucial question then becomes: Which
classes of AGN have jets? In Falcke (1994) and Falcke \& Biermann (1995) we
wrote down a bold (or should one say naive?) hypothesis, simply
stating that since black holes do not have many free parameters, AGN
should be similar in their basic properties (``the universal engine'',
Falcke 1996a) and hence one should {\it ab initio} assume that all AGN have
relativistic jets rather than only a few sub-classes. As it turned out, this
hypothesis, in its simplicity, was surprisingly successful. Here we
want to discuss the evidence we have recently gathered for the
importance of relativistic jets in other classes of AGN, namely
radio-weak quasars and LINER galaxies.

\titlea{2}{Relativistic jets in radio-quiet quasars}
If one looks at the distribution of the radio-to-optical flux ratios
($R$-parameter) of an optically selected quasar sample (here the PG
quasar sample) one finds a clear dichotomy between radio-loud and
radio-quiet sources. This is especially true if one only selects
steep-spectrum quasars, which are supposedly unaffected by orientation
effects. VLA observations of the steep-spectrum radio-loud PG quasars
(Miller, Rawlings, \& Saunders 1993) and Kellerman et al. (1994) have
clearly established, that those sources have FR$\,$II-type radio jets.
This dichotomy was occasionally attributed to the fact that
radio-quiet quasars do not show and do not have radio-jets at
all. However, as we all know, `absence of evidence is not evidence of
absence' --- especially not, if one has not even looked yet.

\titleb{2.1}{Predictions for boosted radio-quiet jets}
Let us therefore ask: what would be the consequences, if radio-quiet
quasars too would have relativistic jets? As for radio-loud quasars,
the most prominent sources would be those which are pointing towards
us and are relativistically boosted. In an optically selected sample, we
would expect that, if radio-quiet quasars have relativistic jets, some
of the quasars are accidentally pointing towards us, thus producing a
population of `weak blazars' with the following properties:

\item{a)}  similar to flat-spectrum, core-dominated, variable radio quasars but with relative low $R$-parameter,

\item{b)} brightness temperatures close to $\sim10^{12}$K or above,

\item{c)} superluminal motion,

\item{d)} very faint (i.e.~radio-quiet) extended radio emission,

\item{e)} number of sources in a well selected sample, and their
Doppler-boosting relative to radio-quiet quasars both imply the same
Lorentz factor,

\item{f)} luminosity- and $z$-distribution consistent with radio-quiet
parent population,

\item{g)} host galaxies compatible with those of radio-quiet quasars.

This list is quite helpful, as it allows an either/or decision: if we
do not find a population of weak blazars, we can exclude that
relativistic jets in radio-quiet quasars exist (or one would have to
invent an argument why those jets never point towards us); if we find
them, we can prove that radio-quiet quasars must have relativistic
jets. Interestingly, in the PG quasar sample, we indeed find a
population of quasars, which at least partially fulfill most of the
criteria listed above and most likely are such weak blazars.

\titleb{2.2}{Radio-intermediate quasars}
Miller et al. (1993) and Falcke et al. (1995 \& 1996a) identified a
small sample of radio-intermediate quasars (RIQ) which sparsely fill
the space in $R$ between radio-loud and radio-quiet quasars. They have
optical+UV luminosities between $10^{45}$ and $10^{47}$ erg/sec, just
like the bulk of the radio-quiet quasars, and unlike radio-loud
quasars which can be found only above $10^{46}$ erg/sec in the PG
sample. They are typical flat-spectrum, core-dominated quasars, but
their $R$ parameter is too low for them to be boosted radio-loud
quasars. Their number and $R$-distribution compared to the radio-quiet
quasars would indicate a bulk Lorentz factor of 2-4.  For at least the
three low-redshift RIQ, there is no extended emission above a level of
a few mJy---far below what is expected for any radio-loud
quasar---neither on the VLA A- \& D-array (Kellerman et al. 1994) nor
on the EVN \& MERLIN scales (Falcke et al. 1996b). At least one source,
III Zw~2, has shown outbursts, indicating a brightness temperature of
$10^{12}$ K (Ter\"asranta \& Valtaoja 1994) which requires relativistic
boosting, while VLBI observations of the three low-$z$ sources
indicate at least lower limits of several $10^{10}$ K. Hence, those
sources meet all the requirements for intrinsically radio-quiet
quasars, whose relativistic jets accidentally point towards us.

\titleb{2.3}{Host galaxies}
In the meantime, since the papers have been published, one other
prediction has been verified. In Falcke et al.~(1996b), we suggested
that in order to test the idea of the RIQ being intrinsically
radio-quiet, at least half of the flat-spectrum RIQ should have spiral
host galaxies. So far, powerful radio galaxies and radio-loud quasars
have turned out to reside in elliptical hosts, while radio-quiet
quasars seem to reside in a mix of spiral and elliptical galaxies
(Kukula et al.~1997). Luckily, two of the three low-redshift RIQ were
part of recent host galaxy studies: HST observations of PG
1309+355 (Bahcall et al.~1997) and NIR observations of III~Zw~2
(Taylor et al.~1996) have now shown that indeed both galaxies are 
spirals. This finally confirms, that the RIQ cannot be and never will
be radio-loud quasars (as they have been classified occasionally in
the past).

\titleb{2.4}{Direct observations of jets}
The only piece missing is direct confirmation of relativistic jets in
radio-quiet quasars; specifically superluminal motion has not yet
been observed. This is, however, not surprising given the
observational difficulties for these weak sources. VLBI
observations of radio-quiet quasars have just recently begun and even
they lack the sensitivity to detect additional components besides the
core. Deep, long-integration VLBI observations of radio-quiet quasars
are certainly needed. On the other hand we can at least give a
preliminary answer to the question whether direct evidence for jets in
radio-quiet quasars exists at all. VLA
observations of Kellerman et al.~(1994) have already revealed a number of
radio-quiet quasars with weak, bi-polar radio-structure. Moreover,
there is a certain regime, where the Seyfert and quasar classifications blend
into each other. In Fig.~1 we show a VLA map of Mrk~34
(Falcke et al. 1997) which is classified as a Seyfert 2 galaxy, but
has an [O {III}] luminosity which is typical for a radio-quiet PG quasar with
an optical+UV luminosity of several $10^{45}$ erg/sec.  The initial
snapshot map of this galaxy (Ulvestad \& Wilson 1984) also looked like some of the
structures Kellerman et al.~(1994) found in some PG quasars. The map in Fig.~1
now shows what kind of radio-quiet (!) jets one can get, if one
integrates long enough.

\begfig 4 cm 
\vskip40pt\hskip110pt
\includegraphics{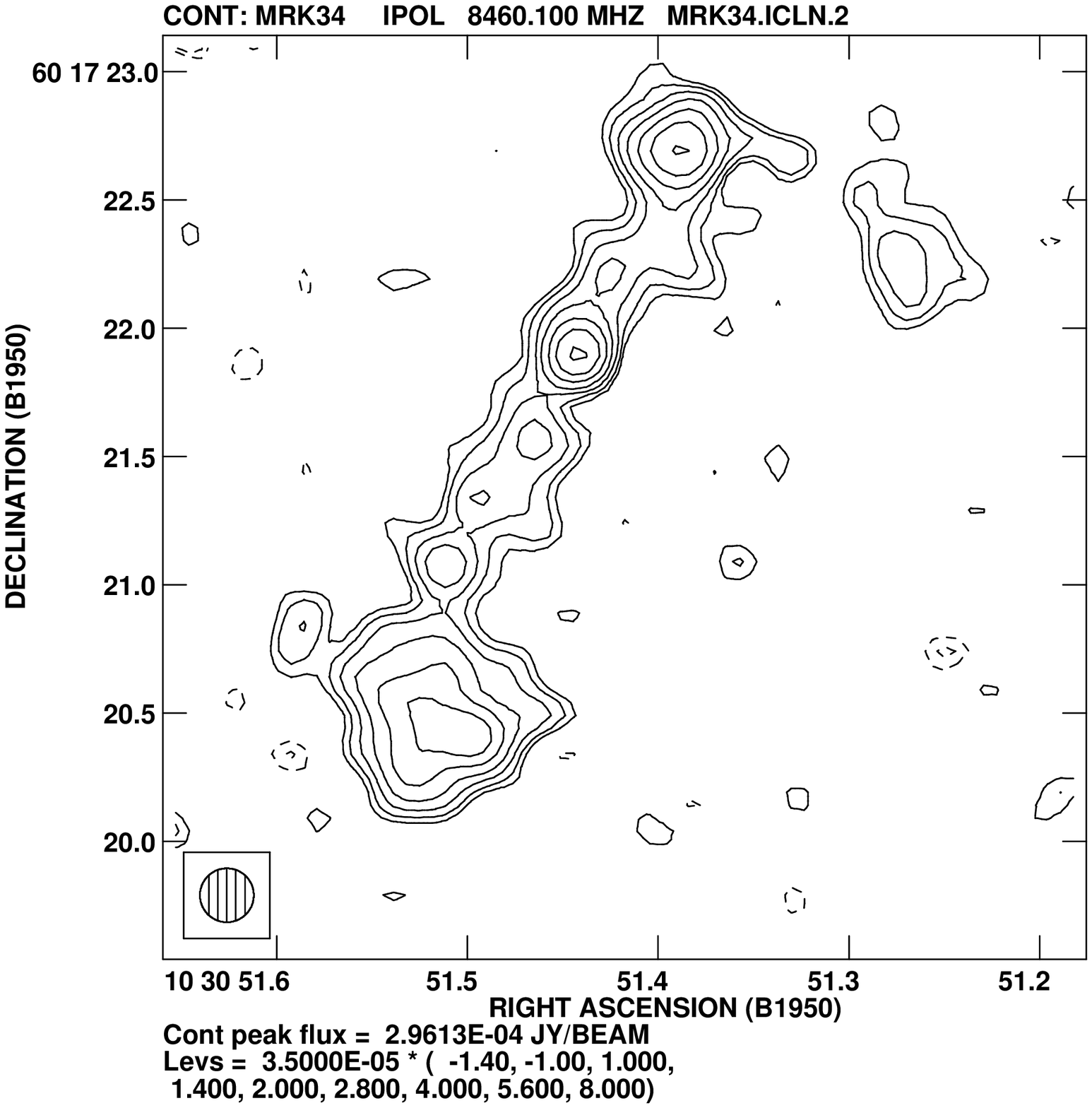}
\vskip-40pt
\figure{1}{VLA map at 3.5cm of Mrk~34, a `Seyfert 2' galaxy with an 
[O {III}] luminosity typical for a radio-quiet quasar (from Falcke
et al.~1997); the beam is 0.25$^{\prime\prime}$. Only long integration
reveals the beautiful jet structure in this galaxy.}
\endfig

\titlea{3.}{Relativistic jets in LINER galaxies}
Besides radio-quiet quasars, there is another regime where one might
be able to find relativistic jets. What happens with those jets in
quasars if the accretion rate becomes lower and lower? Will those jets
die completely, implying that accretion near the Eddington limit is
required for the jet formation, or will the jet just become
proportionally weaker, implying that jet formation is an integral part
of accretion physics? To learn more about this question one first has
to search for and then study jets in low-luminosity AGN. Ho et
al.~(1995 \& 1997) found that roughly one half of nearby galaxies show
signs of nuclear activity, in the form of LINER or Seyfert
spectra. The bolometric luminosities of these AGN (excluding the host
galaxy of course) are in the range $10^{41}-10^{44}$ erg/sec. Heckman
(1980) has speculated that LINER galaxies may preferentially host
compact radio cores in their nuclei; these cores could be interpreted
as scaled down versions of the compact radio cores and jets in
radio-loud quasars (Falcke 1996b\&c).

To test this, we have performed a VLA A-array survey at 2cm of 48
nearby LINER galaxies (Falcke, Wilson, \& Ho in prep.) from the Ho et
al.~(1995) sample to search for compact, flat-spectrum radio
nuclei. The $5\sigma$ detection limit of the survey was $1$ mJy. In
total we detected 21 galaxies at this wavelength. Eleven of them have
flat spectra, six have steep spectra, and four have, as yet, undetermined
spectra. Further VLA observations at 6cm and 3.5cm of these sources are
currently being reduced, so that in the near future we expect to have
complete spectral information for all galaxies. We note that out of
the 11 flat-spectrum sources, 9 are in spiral galaxies.

Our detection rate of flat-spectrum, compact nuclei at 2cm is
relatively high and confirms the initial hunch that LINERs would make
a good sample to detect compact radio nuclei. For comparison, Vila et
al. (1990) looked at a sample of Sbc galaxies with nuclear radio
components and only detected 2 flat-spectrum nuclei in a sample of 27
galaxies---both of those galaxies were LINERs. In elliptical galaxies,
however, the detection rate of compact nuclei is higher (Wrobel \&
Heeschen 1991).

\begfig 4 cm 
\vskip60pt\hskip83pt
\includegraphics{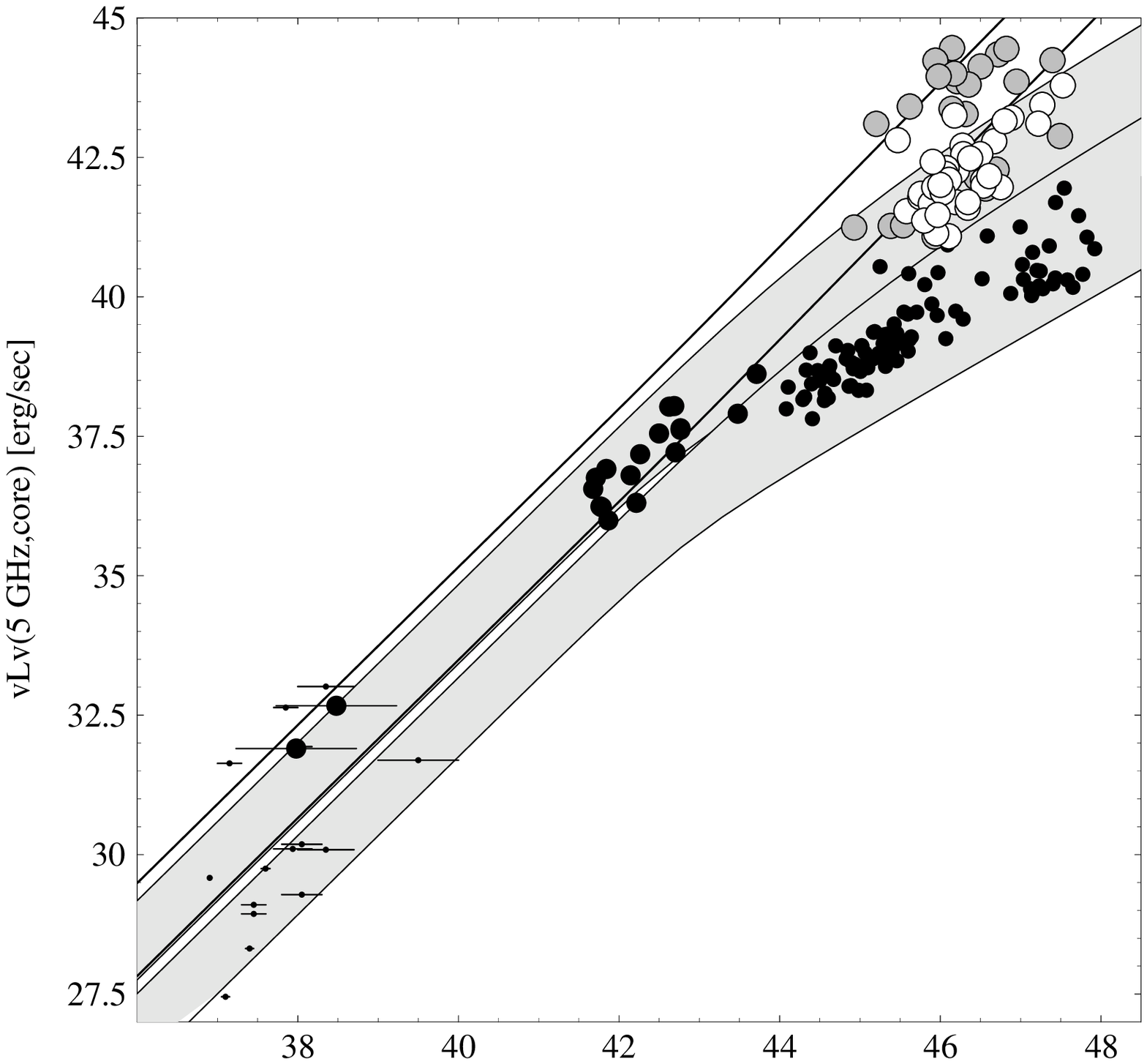}
\vskip-60pt\vskip20pt
\centerline{log $L_{\rm disk}$ [erg/sec]}
\vskip-20pt
\figure{2}{Correlation between accretion disk luminosity (i.e. nuclear
optical+UV luminosity for AGN) and monochromatic radio core luminosity
at 5 GHz. The shaded bands are the theoretical predictions as
presented in Falcke \& Biermann (1996) of radio core luminosities as a
function of accretion disk luminosity for relativistic jets with
randomly oriented inclination angles. The radio cores of the newly
added LINER galaxies are given by big dots, quasars are given by open
(steep-spectrum, radio-loud) and filled circles (flat-spectrum,
radio-loud), and smaller dots (radio-quiet) above $L_{\rm
disk}=10^{44}$ erg/sec. Sources in the lower left are Sgr A*, M31* and stellar
mass black holes (see Falcke \& Biermann for more details).}
\endfig

Of course, the mere fact that we find these radio nuclei in LINERs
does not prove yet that those radio cores are indeed related to the
active nucleus or that they are jets. We therefore looked at the
relation between radio and optical H$\alpha$ flux in those galaxies
and found a good correlation. This is in line with earlier claims of a
connection between optical and radio activity (Ekers \& Ekers 1973,
O'Connell \& Dressel 1978).

Hence, we conclude that the radio cores in LINERs are indeed part of
the central engine. Moreover, we can compare the radio and
emission-line luminosities with the jet/disk model by Falcke \&
Biermann (1996), to learn more about their nature.  The model
predicted a specific radio/nuclear luminosity correlation for
low-power AGN and is based on the assumption that accretion disk
luminosity and jet power in AGN are coupled by a universal
constant. We note that this may remain true even for advection
dominated disks, as have been discussed for LINERs (Lasota et
al. 1996), if the radiative efficieny of the radio and the optical emission in
a jet/disk system are reduced in the same way (i.e. the jet power is
proportional to the energy dissipated in the disk rather than to the
accretion rate).

For a randomly selected (and randomly oriented) sample, the width
(`scatter') of the radio-to-nuclear UV distribution is given by the
typical Lorentz factor of the jets. In Fig.~2 we reproduce (without
changing any parameters) the figure from Falcke \& Biermann, where the
model prediction for low-power AGN was given as shaded bands for a jet
Lorentz factor of $\sim2$. We then converted the {\it narrow}
H$\alpha$ line-luminosities of the LINERs with detected flat-spectrum
nuclei to optical+UV luminosities, using the same proportionality
factors as for the quasars\footnote{$^1$}{See Falcke et al.~1995 \& Falcke
1996a. The exact conversion factors for LINERs require of course a
more thorough discussion. For a few examples (e.g. M81, NGC 4252) this
method at least seems to give a reasonable estimate for the nuclear
luminosity}, and inserted them into the correlation (big dots).
Obviously the LINERs fall exactly into the range predicted for
low-luminosity, {\it radio-loud} jets. This confirms a preliminary
version of this diagram which was presented in Falcke (1996c), but was
based only on a few ill-selected, famous LINER galaxies.

This result not only strongly suggests that LINERs do have powerful
nuclear radio jets --- for some individual cases this is know already
(e.g. M87; NGC4258, Herrnstein et al. 1997; M81, Bietenholz et
al. 1996, etc.) --- but is also consistent with mildly relativistic
Lorentz factors around $\gamma_{\rm j}\simeq2$ as used in the
model. That should be compared with Lorentz factors of
$\gamma\simeq6-10$ derived with the same method for radio-loud quasars
(Falcke et al. 1995).  For the lower Lorentz factor in LINERs (and
also in the Galactic superluminal sources) one can give a very simple
explanation, since this terminal velocity is naturally obtained by a
relativistic plasma in a simple pressure driven jet (Falcke 1996b). To
explain the velocities in radio-loud and radio-quiet quasars, however,
one needs an extra mechanism that provides the additional push
necessary to go beyond $\gamma_{\rm j}=3$.

\titlea{4}{Conclusions}
Based on the experience gathered on relativistic jets in quasars and
radio galaxies, one can make a number of specific predictions for
signs of relativistic jets in other samples of AGN. Especially for
radio-weak quasars a number of those predictions has been verified for
a sample as well as for individual cases, making it very likely that
relativistic jets do in fact exist in many, if not all, radio-quiet
quasars. A search for similar jets in low-power galaxies has just
begun, but there is already data suggestive of the existence of
relativistic jets in LINERs. The study of those sources will greatly
expand our horizon and may eventually help us to understand the
underlying principle governing the formation of jets in general.

\bigskip\noindent
{\it Acknowledgment:} We thank Joan Wrobel and Jim Ulvestad for help
during the VLA data reduction. This research was supported by NASA
under grants NAGW-3268, NAGW4700, and NAG8-1027

\begrefchapter{References}
\ref Bahcall, J.N., Kirhakos, S., Saxe, D.H., Schneider, D.P. 1997, ApJ 479, 642
\ref Bietenholz, M.F., Bartel, N., Rupen, M.P., et al. 1996, ApJ 604, 28
\ref Ekers, R.D., \& Ekers, J.A. 1973, A\&A 24, 247
\ref Falcke, H. 1994, Dissertation, RFW Universit\"at Bonn
\ref Falcke, H. 1996a, in: ``Jets from Stars and Galactic Nuclei'',
Lecture Notes in Physics 471, W. Kundt (ed.), Springer, p. 19--34 
\ref Falcke, H. 1996b, ApJ 464, L67
\ref Falcke, H. 1996c, in: ``The Galactic Center'', ASP Conf.~Ser.~102, R. Gredel (ed.), 453--461
\ref Falcke, H., Biermann P.L. 1995, A\&A 293, 665 
\ref Falcke, H., Biermann P.L. 1996, A\&A 308, 321
\ref Falcke, H., Malkan M., Biermann P.L. 1995, A\&A 298, 375 
\ref Falcke, H., Sherwood, W., Patnaik, A. 1996a, ApJ 471, 106
\ref Falcke, H., Patnaik, A., Sherwood, W. 1996b, ApJ 473, L13
\ref Falcke, H., Wilson, A.S., Simpson, C. et al. 1997, ApJ, to be submitted
\ref Heckman, T.M. 1980, A\&A 87, 152
\ref Ho, L.C., Filippenko, A.V., Sargent, W.L.W. 1995, ApJS, 98, 477
\ref Kellermann, K.I., Sramek, R., Schmidt, M., et al. 1994, AJ 108, 1163
\ref Kukula, M.J., Dunlop, J.S., Hughes, D.H., Taylor, G., Boroson, T. 1997, in ``Quasar Hosts'', ESO/IAC conference, Tenerife [astro-ph/9701192]
\ref Lasota, J.-P., Abramowicz, M.A., Chen, X., et al. 1996, ApJ 462, 142
\ref{Miller, P., Rawlings, S., \& Saunders, R. 1993, MNRAS 263, 425}
\ref O'Connell, R.W., Dressel, L.L. 1978, Nat 276, 374
\ref Taylor, G.I., Dunlop, J.S., Hughes, D.H., Robson, E.I. 1996, MNRAS 283, 930
\ref Ter\"asranta, H., \& Valtaoja, E. 1994, A\&A 283, 51
\ref{Ulvestad, J.S. \& Wilson, A.S. 1984, ApJ 278, 544}
\ref Vila, M.B., Pedlar, A., Davies, R.D., Hummel, E., Axon, D.J. 1990, MNRAS 242, 379
\ref Wrobel, J.M., \& Heeschen, D.S. 1991, AJ 101, 148
\endref
\bye